# Electrical control of a solid-state flying qubit

Michihisa Yamamoto[1, 2], Shintaro Takada[1], Christopher Bäuerle[1, 3], Kenta Watanabe[1], Andreas D. Wieck[4] & Seigo Tarucha[1, 5]

[1]*Department of Applied Physics, University of Tokyo, Bunkyo-ku, Tokyo, 113-8656, Japan,* [2]*ERATO-JST, Kawaguchi-shi, Saitama 331-0012, Japan,* [3]*Institut Neel – CNRS and Université Joseph Fourier, 38042 Grenoble, France,* [4]*Chair of Solid-State Physics, Ruhr-Universität Bochum, D-44780 Bochum, Germany,* [5]*ICORP (International Cooperative Research Project) Quantum Spin Information Project, Atsugi-shi, Kanagawa, 243-0198, Japan.*

**Solid-state approaches to quantum information technology are attractive because they are scalable. The coherent transport of quantum information over large distances, as required for a practical quantum computer, has been demonstrated by coupling solid-state qubits to photons[1]. As an alternative approach for a spin-based quantum computer, single electrons have also been transferred between distant quantum dots in times faster than their coherence time[2, 3]. However, there have been no demonstrations to date of techniques that can coherently transfer scalable qubits and perform quantum operations on them at the same time. The resulting so-called flying qubits are attractive because they allow for control over qubit separation and non-local entanglement with static gate voltages, which is a significant advantage over other solid-state qubits in confined systems for integration of quantum circuits. Here we report the transport and manipulation of qubits over distances of 6 microns within 40 ps, in an Aharonov-Bohm ring connected to two-channel wires that have a tunable tunnel coupling between channels. The flying qubit state is defined by the presence of a travelling electron**

**in either channel of the wire, and can be controlled without a magnetic field. Our device has shorter quantum gates, longer coherence lengths (~86 μm at 70 mK), and shorter operation times (~10 ps or 100 GHz) than other solid-state flying qubit implementations[4, 5], which makes our solid-state flying qubit potentially scalable.**

The flying qubit device which we present in this letter is an electrical analog to an optical two-path interferometer, in which the qubit basis is defined by the presence of a travelling electron in either of the two paths. Similar electrical two-path interference has been achieved in Mach-Zehnder interferometers of quantum hall edge channels[4, 5], which have been the testing grounds for quantum physics like orbital entanglement[6, 7]. However, these devices exhibit edge channel coherence lengths of 24 μm at 20 mK[5]. This, together with a requirement for high magnetic fields, limits their scalability.

Relative to edge-channel two-path interferometers, our two-path interferometer operates at lower magnetic field, has well-defined quantum operations, shorter quantum gates, and longer coherence lengths. Despite its apparent simplicity, realization of such an electrical two-path interferometer at low magnetic fields is highly challenging. The main difficulty comes from the nature of electrons on the Fermi surface: electrons usually take multiple paths, so that the quantum phase of an electron is easily scrambled. This causes the quantum information being carried by the electron to be lost.

Electrons travelling through an Aharonov-Bohm (AB) interferometer[8-11], generally known as a two-path interferometer, also suffer from the existence of multiple paths. Electron interference arises in AB interferometers due to the phase the electron acquires in one path relative to the other, with the phase difference given by $\Delta\varphi = \oint \mathbf{k} \cdot d\mathbf{l} + \frac{e}{\hbar} BS$. Here, $B$ is the perpendicular magnetic field and $S$ is the area enclosed by the AB ring. This phase difference causes an oscillation of the current as a function of *B* with a period of $h/eS$. It turns out, however, that the two-terminal linear conductance through an AB ring suffers from so-called phase rigidity[12-14]. Onsager's law for linear conductance[12, 15, 16], $G(B) = G(-B)$, implies that the phase of the AB oscillation can take only the values 0 or π at *B* = 0. To satisfy this boundary condition imposed by the contact geometry while still allowing for a continuous phase variation *Δφ*, contributions from multiple scattered paths needs to be added to *Δφ*. What is usually observed in an AB experiment is therefore not an ideal two-path interference, but a complicated multi-path interference. Such conventional AB interference is shown in Figs. 1b and 1c. Multi-terminal devices have also been employed to extract contribution of the two-path interference and avoid phase rigidity[17]. However, the multi-path contribution is still present even in such interferometers, and leads to the loss of quantum information carried by coherent electrons reaching other reservoirs.

Here, we realize a novel two-path interferometer, which can act as a flying qubit, by combining the AB ring with two-channel wires, i.e. parallel tunnel-coupled quantum wires [18, 19] that allow tunnelling of an electron between the two paths (see Fig. 1). We define the two pseudo spin states $|\uparrow\rangle$ and $|\downarrow\rangle$, where $|\uparrow\rangle$ and $|\downarrow\rangle$ correspond to the state having an electron in the upper and lower path, respectively[20-22]. In such a structure, any superposition state of $|\uparrow\rangle$ and $|\downarrow\rangle$ in the ring can transmit into the tunnel-coupled wire by being directly transformed into the superposition of the bonding and anti-bonding state in the tunnel-coupled wire, i.e. $\psi_S = \frac{1}{\sqrt{2}}(|\uparrow\rangle + |\downarrow\rangle)$ and $\psi_{AS} = \frac{1}{\sqrt{2}}(|\uparrow\rangle - |\downarrow\rangle)$. This is in contrast with the conventional AB ring with only single wire leads, where only $\psi_S$ is transmitted into the leads. Consequently, this new structure works as a two-path interferometer for ballistic electrons, which does not suffer from paths encircling the AB ring due to the absence of backscattering at the entrance of the tunnel-coupled wire. The pseudo spin is then defined as a flying qubit. It is important to mention that even though the electron wave spreads over the interferometer under a low energy excitation, the pseudo spin state is explicitly defined at each position in the interferometer (see supplementary information). Therefore, a flying qubit state carried by each single electron is strictly defined as a function of the position without uncertainty.

Our device (Fig. 1) was defined by surface Schottky gates in a two-dimensional electron gas at the interface of a high mobility GaAs/AlGaAs heterostructure. The negative gate voltages $V_{T1}$ and $V_{T2}$ applied on the orange-coloured central gates allow control of the tunnel coupling between the upper and lower paths. Since these gates are made as narrow as 50 nm at the tunnel-coupled regions, we can keep the tunnel coupling at the right coupled-wire strong enough even when the central region of the device is depleted with $V_{T1}$ to form an AB ring. Side gates of the AB ring, $V_{M1}$ and $V_{M2}$ allow to modulate the Fermi wave vectors in the ring. All measurements except for the ones in Figs. 2 and 4 were carried out in a dilution refrigerator at 70 mK.

The initial qubit state is simply defined as $|\Uparrow\rangle$ by injecting electrons into only one of the two wires. The projection of the final state is obtained by measuring the output currents, $I_\Uparrow$ and $I_\Downarrow$. The tunnel coupling between the two wires yields the hybridized symmetric and anti-symmetric basis states, $\psi_S$ and $\psi_{AS}$ with confinement energy $E_S$ and $E_{AS}$, respectively. The energy spacing $\Delta E = E_{AS} - E_S$ is given by the tunnel coupling energy. The quantum state of an electron wave, which is a superposition state of $\psi_S$ and $\psi_{AS}$, then acquires a relative phase $\theta = \frac{1}{\hbar}\Delta E \tau$ between $\psi_S$ and $\psi_{AS}$, where $\tau \approx L_t/v_F$ is the traversal time for electrons propagating over the length $L_t$ of the tunnel coupling with the Fermi velocity $v_F$. This $\theta$ is the rotation angle of a state vector about the x-axis on the Bloch sphere (see supplementary information). This rotational operator is expressed as a matrix,

$$R_x(\theta) = \begin{pmatrix} \cos\dfrac{\theta}{2} & i\sin\dfrac{\theta}{2} \\ i\sin\dfrac{\theta}{2} & \cos\dfrac{\theta}{2} \end{pmatrix}.$$

It follows, that if an electron is injected into one of the wires, the pseudo spin state at the end of the tunnel-coupled wire periodically oscillates between $|\uparrow\rangle$ and $|\downarrow\rangle$ as a function of the tunnel coupling energy due to this rotational operation.

We demonstrated $R_x(\theta)$ by depleting the region beneath the gate $V_{T2}$. A current is injected into one of the wires and measured via output currents $I_\uparrow$ and $I_\downarrow$ as a function of the gate voltage $V_{T1}$. Evolution of the pseudo spin state of an electron propagating through the tunnel-coupled wire is schematically depicted in Fig. 2a, where the rotation angle $\theta$ depends on $V_{T1}$. Clear anti-phase oscillations of the currents $I_\uparrow$ and $I_\downarrow$ were observed at 2.2 K as shown in Figs. 2b and 2c, which cancel each other in the total current $I_\uparrow + I_\downarrow$. The total current has only weak dependence on $V_{T1}$. This means that there is basically no backscattering for the oscillating signal transmitted through the tunnel-coupled wire. Therefore, the anti-phase oscillations evidence the oscillation of the electron between the two wires induced by the tunnel coupling. Here the visibility of the oscillation, the ratio of the oscillation component to the total current, is limited to about a percent. This is due to the existence of a few transmitting channels and high measurement temperature (see also supplementary information). Note that at low temperature, disorder scattering in the quantum wires leads to complicated fluctuations of the output currents as a function of $V_{T1}$, which masks the current oscillation induced by the tunnel coupling.

A rotation about the z-axis can be achieved in the AB ring by varying the magnetic field or the gate voltages $V_{M1}$ and $V_{M2}$. The relative transmission phase $\varphi$ between $|\uparrow\rangle$ and $|\downarrow\rangle$ is given by

$$\varphi = \int_P^Q \left(\mathbf{k} - \frac{e}{\hbar}\mathbf{A}\right) d\mathbf{l}_\downarrow - \int_P^Q \left(\mathbf{k} - \frac{e}{\hbar}\mathbf{A}\right) d\mathbf{l}_\uparrow = \oint \mathbf{k} \cdot d\mathbf{l} - \frac{e}{\hbar}BS$$

with P (Q) the position where the wave splits (recombines), and is defined as a rotation angle for the rotational operator about the z-axis on the Bloch sphere,

$$R_z(\varphi) = \begin{pmatrix} 1 & 0 \\ 0 & e^{i\varphi} \end{pmatrix}.$$

The combination of $R_x$ and $R_z$ enables the generation of an arbitrary vector state on the Bloch sphere. Arbitrary rotation of the state is also possible by simultaneously controlling the tunnel coupling and the difference of the transmission phase between the two paths.

$R_z(\varphi)$ was demonstrated in the scheme of a Ramsey interference (see Fig. 3a) using a high velocity channel which is insensitive to the disorder. We established the two sets of the tunnel-coupled wires to prepare $R_x(\pi/2)$ (see supplementary information) and varied the magnetic field to prepare $R_z(\varphi)$ in the AB ring. The measured $I_\uparrow$ and $I_\downarrow$ plotted in Fig. 3b show AB oscillations with exactly opposite phases, indicating that the two-path interference is dominant over multi-path contributions of backscattering. The difference in amplitude between $I_\uparrow$ and $I_\downarrow$ is the backscattered term and is much smaller compared to the main oscillation.

To further confirm the two-path interference and the electrical control of the qubit, we also swept the gate voltages $V_{M1}$ and $V_{M2}$ to modulate the Fermi wave vector $k_F$ of the respective paths as shown in Figs. 3c and 3d. The phase evolves almost linearly and smoothly over $2\pi$ with the gate voltages in strong contrast to Fig. 1c. These results hence demonstrate ideal two-path interference with absence of phase rigidity and that the qubit can be controlled even without magnetic field.

The flying qubit presented here is promising for quantum information technology. In addition to the ability to transfer the quantum information over a long distance, it has a much shorter operation time compared to other qubits in solid-state systems. The operation time $L/v_F$ ($L$: gate length, $v_F$: Fermi velocity) is of the order of 10 ps (see supplementary information). We also found by analysing the temperature dependence of the oscillation amplitude (Fig. 4), that this qubit has a very long coherence length $l_\varphi$. Assuming that the electron travels with constant velocity, exponential decay of the amplitude[23-26] as observed in Fig. 4 leads to a relationship of $l_\varphi \propto 1/T$. We obtained $l_\varphi$ = 86 μm at 70 mK from the decaying rate (see Fig. 4 caption). This value is much larger than the one reported in the Mach-Zehnder interferometer of an edge state[5] in which limited coupling between the interferometer and the reservoir via only a few edge channels makes the interferometer sensitive to the charge fluctuation[23].

The only drawback at present is the limited visibility defined as the AB oscillation amplitude divided by the total current, which is only about 0.3% for the measurements of Figs. 3b-d (see supplementary information). Because the coherence length is found to be much longer than the interferometer length, decoherence is not the origin of the low visibility. Influence of the thermal smearing due to the difference of the Fermi velocity between the two paths is also small at our measurement temperature (see supplementary information). The visibility should be improved by optimizing the device design. There are two possible "geometrical" reasons for the low visibility. Firstly, entrance and exit of the tunnel-coupled wires are made so narrow that the conductance may be restricted there and backscattering may occur. Secondly, there is a contribution from several transmitting channels in each part of the tunnel-coupled wires and in each arm of the AB ring, although only one in each wire contributes to the main oscillation. A possible remedy is to employ a high electron density wafer to define strongly confined single mode channels, and to make the electrons propagate through them without suffering from backscattering. We indeed significantly improved the visibility to about 10% by employing a high electron density wafer and defining narrower channels (see supplementary information).

In addition to the quantum information transfer, it should also be possible to create a non-local entanglement state following the scheme proposed in Refs. 21, 22, combined with single electron sources[27, 28] to synchronize qubits. This flying qubit can also be employed in combination with a static qubit[29]. Employing a high mobility wafer with a mean free path exceeding 100 μm, we would in principle be able to integrate 100 qubits because each quantum operation, including two-qubit operation, is performed within a 1 μm scale (see supplementary information).

**Acknowledgements**

We acknowledge B. Halperin for discussion. M.Y. acknowledges financial support by Grant-in-Aid for Young Scientists A (no. 20684011). S. Takada acknowledges support form JSPS Research Fellowships for Young Scientists. S.Tarucha acknowledges financial support by Grant-in-Aid for Scientific Research S (no. 19104007), B (no. 18340081), MEXT Project for Developing Innovation Systems, MEXT KAKENHHI "Quantum Cybernetics" and JST Strategic International Cooperative Program. A.D.W. acknowledges expert help from PD Dr. Dirk Reuter and support of the DFG SPP1285 and the BMBF QuaHLRep 01BQ1035. C.B. acknowledges financial support from CNRS (DREI) - JSPS (no. PRC 424 and L08519).

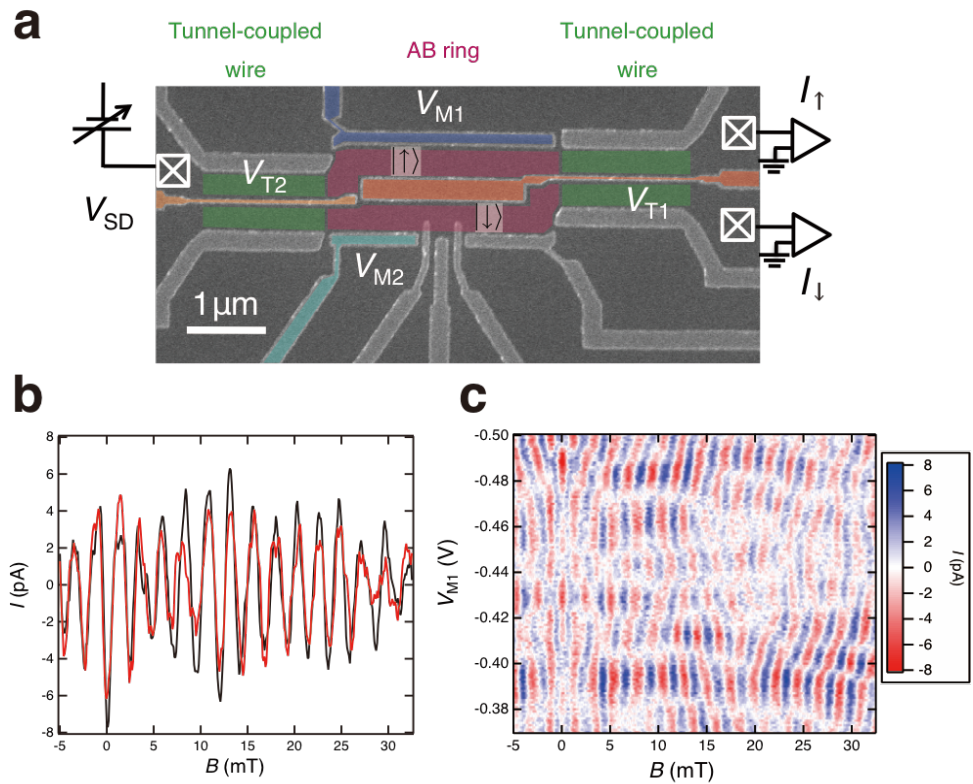

**Figure 1 Device image and observed conventional Aharonov-Bohm oscillation. (a)** SEM image of the flying qubit device and schematic of the experimental setup. The device was defined by Schottky gates in an n-AlGaAs/GaAs 2DEG based heterostructure (2DEG: n = $1.9\times10^{11}$ cm$^{-2}$, μ ~ $2\times10^6$ cm$^2$/Vs, depth = 125 nm) using standard split-gate techniques. The mean free path is 14.5 μm. $V_{T1}$ and $V_{T2}$ allow for control of the tunnel coupling between the parallel quantum wires and $V_{M1}$ and $V_{M2}$ modify the Fermi wave vector at each path. Three cross-in-square symbols represent ohmic contacts. **(b)** Typical AB oscillation measured at T = 70 mK for $V_{T2}$ = 0 and $V_{M1}$ = -0.391 V. The black and red curves are the currents $I_\uparrow$ and $I_\downarrow$ obtained at the upper and the lower contacts, respectively. The negative gate voltage $V_{T1}$ is also set to be low enough ($V_{T1}$ = -0.300 V) so that the propagating electrons do not occupy the anti-symmetric orbital in the right tunnel-coupled wire. Then this three-terminal

device works effectively like a two-terminal device, resulting in a similar oscillation between $I_\uparrow$ and $I_\downarrow$. The oscillation components are extracted by subtracting a smoothed background. **(c)** Intensity plot of $I_\uparrow$ as a function of the perpendicular magnetic field and $V_{M1}$. Due to phase rigidity, the phase at *B*=0 is either 0 or π. Sweeping $V_{M1}$, many π jumps are observed in the phase, which occur due to the multi-path interference.

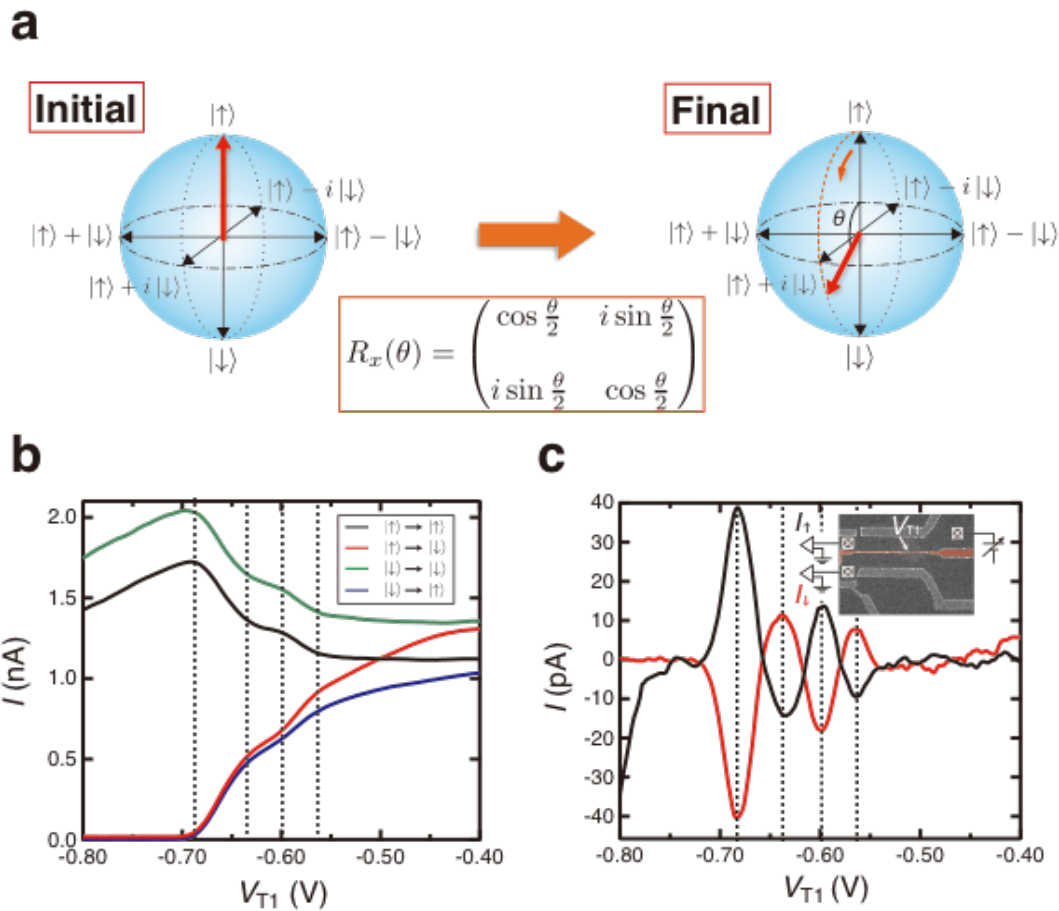

**Figure 2 Demonstration of $R_x$ operation. (a)** Evolution of the flying charge qubit state via the tunnel coupling. **(b)** Measured output currents in the tunnel-coupled wire. There are at least 3 transmitting modes in each wire (supplementary information). Black and red curves are the output currents $I_\uparrow$ and $I_\downarrow$ measured as a function of the gate voltage $V_{T1}$ for $V_{sd}$ = 50 µV, $V_{T2}$ = -0.84 V and $T$ = 2.2 K when the current is injected from the upper wire. Blue and green curves are the measured output currents $I_\uparrow$ and $I_\downarrow$ when the current is injected to the lower wire with the same gate voltage configuration. Coincidences between the black and the green curves and between the red and the green curves indicate that the oscillation of the current is due to the tunnel coupling

between the two paths. Note that the measurement of the current at such a high *T* does not suffer from universal conductance fluctuations originating from disorder. **(c)** Oscillating component of the output currents. Smoothed backgrounds are subtracted from the black and red curves of Fig. 2b.

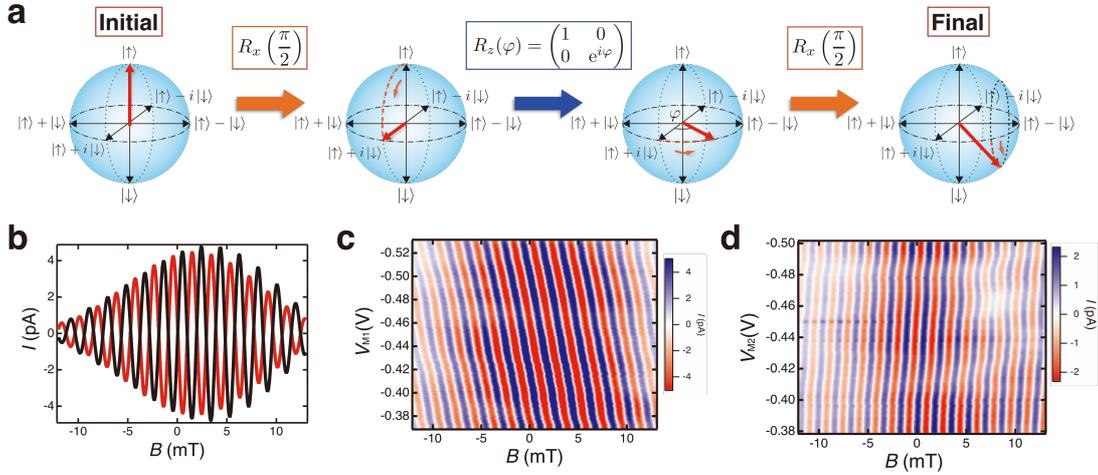

**Figure 3 Demonstration of $R_z$ operation. (a)** Evolution of the flying qubit state in the scheme of Ramsey interference. **(b)** Output current oscillation as a function of the magnetic field. Black and red curves are the output current $I_\uparrow$ and $I_\downarrow$, respectively, measured for $V_{sd}$ = 50 µV. Both tunnel-coupled wires work as $R_x(\pi/2)$, which results in maximum current oscillation (see supplementary information). **(c)** Intensity plot of $I_\uparrow$ as a function of the magnetic field $B$ and the side gate voltage $V_{M1}$. **(d)** Intensity plot of $I_\uparrow$ as a function of $B$ and $V_{M2}$. The oscillation components are extracted by a Hamming window followed by a FFT filter for Figs. 3b-d. In both Figs. 3c and 3d, there is no phase jump of π in contrast to Fig. 1c, which is a proof of the two-path interference. The smooth phase shift is given by $\Delta\varphi = \Delta k_F \cdot L_i = \dfrac{\pi C_i L_i}{2e} \Delta V_{Mi}$, where $L_{1(2)}$ is the length of the upper (lower) path and $C_{1(2)}$ is the capacitance per length between the gate and the upper (lower) quantum wire. We obtained $C_1 \approx C_2 \approx$ 2.9 pF/m from the data in Figs. 3c and 3d.

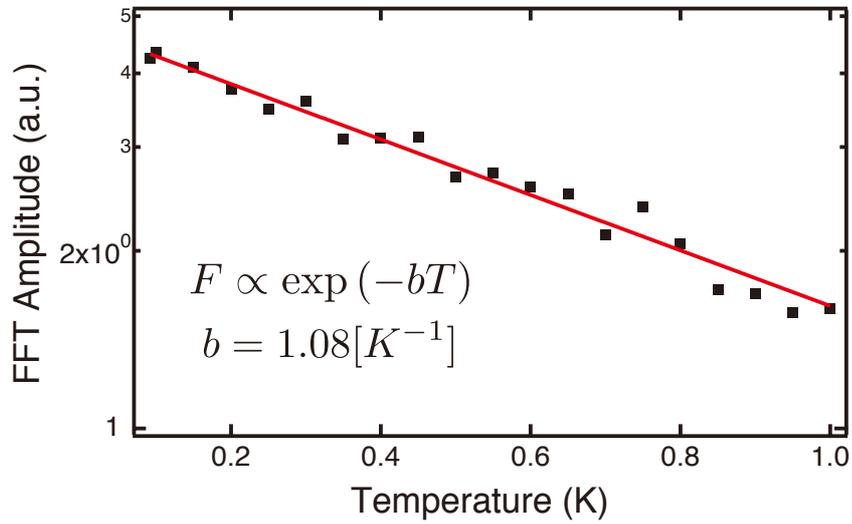

**Figure 4 Evaluation of coherence length $l_\varphi$.** Temperature dependence of the FFT amplitude, $F$. The exponential decay suggests that the decoherence rate is proportional to $T$ and therefore $l_\varphi \propto 1/T$ assuming $v_F$ is constant. We obtained the slope of 1.08 K$^{-1}$ by fitting the data (red line). Considering that the length of the interferometer is 6.5 µm, we set $l_\varphi$ = 6.5 µm at $T$ = 1/1.08K$^{-1}$ = 0.93 K. With $l_\varphi \propto 1/T$, we estimate $l_\varphi$ ~ 86 µm at $T$ = 70 mK.